\documentclass[12pt, prd]{revtex4}
\usepackage{graphicx,epsfig}
\input{epsf.tex}
\usepackage{amsmath}
\usepackage{amssymb}

\usepackage[utf8]{inputenc}

\begin{document}

\title{Einstein and Eddington and the consequences of general
relativity: Black holes and gravitational waves}

\author{\vskip 0.3cm
Jos\'{e} P. S. Lemos, Carlos A. R. Herdeiro, and Vítor Cardoso}
\affiliation{Centro de Astrof\'isica e Gravita\c{c}\~ao - CENTRA,
Departamento de F\'isica, 
Instituto Superior T\'{e}cnico - IST,
Universidade de Lisboa - UL,
Avenida Rovisco Pais 1, 1049-001, Lisboa, Portugal.\footnote{emails:
joselemos@tecnico.ulisboa.pt, carlosherdeiro@tecnico.ulisboa.pt,
vitor.cardoso@tecnico.ulisboa.pt}}



\begin{abstract}
\vskip 1cm

For the celebrations of the 100 years of the observations undertaken
by Eddington at the island of Principe and collaborators at
Sobral during a total solar eclipse in May 29, 1919, which have
confirmed Einstein's theory of general relativity through the
deflection of the incoming light from distant stars due to the
spacetime curvature caused by the Sun, we highlight the main aspects
of the theory, its tests and applications, focusing on some of its
outstanding consequences. These are black holes, the object par
excellence of general relativity, and gravitational waves, the
gravitational probe for the distant Universe. We also point out some
open issues.

\end{abstract}
\maketitle

\newpage
\section{General relativity}

\subsection{General relativity: a new notion of space, time and matter}
Einstein contributed to many areas of physics: statistical physics,
electrodynamics, solid state physics, quantum mechanics, special
relativity, general relativity, unification theories, foundations of
quantum mechanics and philosophical principles of physics. His
greatest achievement is, undoubtedly, the general theory of
relativity. Max Born, one of the founders of quantum mechanics, wrote
in 1955 ``The foundation of general relativity appeared to me then,
and it still does, the greatest feat of human thinking about Nature,
the most amazing combination of philosophical penetration, physical
intuition and mathematical skill.''  Dirac, famous for having
theoretically discovered the positron, stated in 1968 that general
relativity is ``probably the greatest scientific discovery that was
ever made".  General relativity emerges from the necessity to bring
together Newtonian gravity and the theory of special relativity. Let
us first indicate some notions of both these theories. 

Newtonian gravitation is based on the three laws of Newtonian
mechanics, which rule the motion of particles, together with the
universal gravitation law. On the one hand, the second of the three
laws of Newtonian mechanics specifies the acceleration of a particle
with a certain inertial mass, under the action of a force. The law of
universal gravitation, on the other hand, states that two 
particles, or two bodies,
attract one another with a force which is proportional to their 
gravitational masses
and inversely proportional to the square of the distance between
them. 
Newtonian gravitation had enormous success
in celestial mechanics, describing planetary
and all other type of orbits around the Sun,
as well as in explaining Earth's tides, 
and it is now used to
predict with precision 
the orbits of satellites, amongst many other applications.

Special relativity was formulated in 1905, born out of the need to
make the laws of mechanics compatible with those of Maxwell's
electromagnetism, the theory that describes the electromagnetic
field. Maxwell's theory
established the existence of electromagnetic waves
propagating at the speed of light, and that light is an
electromagnetic phenomenon. 
Key properties of Maxwell's theory were
inconsistent with the laws of Newtonian mechanics. For moving
particles with velocities close to the speed of light, and for light
itself, other laws of mechanics, as well as a different notion of
space and time, were needed. Einstein then postulated that the laws of
physics should be the same for any inertial observer, regardless of
the observer's velocity, and in addition 
that the speed of light should be
the same for all observers. It follows that the speed of light is the
maximal possible velocity for all physical objects, and that space and
time are no longer absolute concepts but rather relative to the
observer. Shortly afterwards, in 1907, the german mathematician
Hermann Minkowski showed that special relativity could be understood
within a unified concept of spacetime, providing a logical and
mathematical framework for the theory. He proposed that spacetime is a
single entity and defined the notion of spacetime interval,
generalizing the Pythagorean distance formula.  Moreover, he
constructed spacetime diagrams, of great use to visualize particle
trajectories. Special relativity has applications in a wide range of
physics fields. The world of particle physics is relativist. The twins
paradox is real: identical twin muons in particle accelerators die,
i.e., decay,
earlier or later according to being at rest or moving,
respectively. While one muon goes around the accelerator
untroubled, when arriving at the starting point its 
twin has long
decayed into an electron.

On the one hand, special relativity shows the maximal velocity for
physical propagation is the speed of light; on the other hand,
Newtonian gravitation propagates instantaneously. It follows that
Newtonian gravitation needed to be framed as a relativistic
theory. Consequently, building upon these two theories, Einstein
formulated a new model that he named general theory of relativity, or
simply general relativity. It is a principle based theory: the
principle of equivalence, the principle of general relativity, the
principle of covariance, the principle of minimal coupling and the
correspondence principle. The most important of these, that guided the
final formulation of the final theory, was the equivalence principle.

An experimental fact, seemingly trivial, drew Einstein's attention:
inertial mass and gravitational mass had experimentally equal
values. This equality implies that all objects, regardless of their
size and composition, when placed in a gravitational field, all fall
with the same acceleration, as had already been observed by Galileo
and Newton. Einstein named this fact the principle of equivalence. As
a profound consequence, this principle implies that locally, inertia
and gravitation are one and the same thing. Indeed, the equivalence
between inertial and gravitational mass implies inertia and gravity
are locally indistinguishable.

In a local region where there is gravitation, one may, by the
equivalence principle, choose an appropriate frame of reference so as
to gauge away locally gravity, making spacetime locally
flat. Gravitation, thus, can be eliminated locally, but not
globally. A non-uniform gravitational field cannot be eliminated
everywhere. In a gravitation-free region, acceleration-free particles
follow straight lines, which are the geodesics of the flat spacetime
of special relativity. In a gravitation-free region, accelerating
particles follow curved lines, but still in the flat spacetime of
special relativity. In a region with gravitation, particles follow
curved lines, the geodesics of a curved spacetime. Einstein
consequently proposed that gravitation is a manifestation of curved
spacetime.

General relativity is, in this way, a geometrical theory of
spacetime. The theory uses Riemannian geometry, which is based on
three concepts. The concept of a metric which measures spacetime
intervals, the concept of a linear connection which defines the
parallel transport of vectors and generic tensors and which is a
function of the metric, and the concept of curvature which in turn is
a function of the connection and thus of the metric. Making a
parallelism with Newtonian gravitation, the metric is the
gravitational potential, the connection is the gravitational force and
the curvature is the tidal force. General relativity geometrizes
gravitation. In the Einstein field equation, spacetime tells matter
how to move and matter tells spacetime how to curve.

\subsection{Tests and major consequences of general relativity}

A single experimental result, namely, the equality between inertial
and gravitational mass, led Einstein to the principle of equivalence,
which, in turn, led him to the creation of general relativity. The
theory was confirmed by experimental tests, it has technological
applications, and led to outstanding developments in our knowledge
about the Universe. Let us consider these issues.

The redshift of light when climbing a gravitational field was
predicted by Einstein in 1907 and confirmed definitely in 1960 by
Pound and Rebka. It is a test of the equivalence principle.  Another
test the theory promptly passed at the same time it was formulated was
the precession of Mercury's perihelion in its orbit around the
Sun. General relativity yields the 43 arc seconds that were missing in
Newtonian gravitation. General relativity also predicts that the
spacetime curvature deflects light rays when they pass close to the
Sun.
This prediction was confirmed by the celebrated expeditions of
Eddington to Principe island and collaborators to Sobral, Brazil, in
May 29, 1919, whose centennial was celebrated this year, 
see https://science.esundy.tecnico.ulisboa.pt/en.  The
observations yielded results compatible with the 1.75 arcseconds
predicted by the theory for a light ray at the Sun's rim. Another test
still is the delay in the radar echo of a signal sent towards a
planet, proposed and confirmed experimentally by Shapiro in 1964,
which again has confirmed the theory.

General relativity also has technological applications. It is
remarkable that for the GPS system to function properly general
relativity must be taken into account. It is the first technological
application of the theory. To compute the position of an object with a
one meter precision, the GPS satellites' clocks must measure time with
a precision of one part in $10^{13}$. For such precision, both special
and general relativity effects must be considered, due to the relative
motion between the Earth and the satellites and the weaker
gravitational field for clocks on the
satellites relatively 
to clocks on the Earth. These effects are of one part in $10^{10}$, and if not
properly considered, the GPS would be useless. This practical
application was certainly not envisaged when the theory was
formulated.  Another application, is the measurement of the snow depth
variations in Mars, with the extraordinary precision of 10 cm, using
laser ranging from satellites orbiting the planet, which requires
similar corrections.

The major and innovative consequences of general relativity are black
holes, gravitational waves, cosmology and unification theories. Black
holes are the objects par excellence of general relativity, made up of
pure gravitation. Their main feature is the existence of an event
horizon, beyond which nothing can be observed by an external observer.
They were predicted in 1939 Oppenheimer and Snyder, two American
physicists.  Gravitational waves are generated by accelerated
interacting masses.  They were finally detected, for the first time on
September 14th 2015. We will discuss black holes and gravitational
waves in more detail below. Cosmology as a physical science, rather
than a metaphysical one, was created with the appearance of general
relativity. The first cosmological model was conceived by Einstein in
1917. In this model the universe is static and space is a
3-sphere. This geometry showed, for the first time, that the Universe
could be finite, but unlimited, i.e., with no boundary. In the
following decade, the Russian physicist Alexander Friedmann and the
Belgium physicist George Lema\^itre both proposed independently models
of an expanding universe, which, after the observational confirmation
by Edwin Hubble in 1929 with the Mount Wilson telescope, are at the
basis of the more sophisticated cosmological models that still prevail
today. Unification theories started to be considered properly after
general relativity.  General relativity is a field theory which
describes the gravitational field as geometry of the spacetime
curvature. Another fundamental field in physics is the electromagnetic
field, described by Mawell's equations. The unification of these two
fundamental fields in a single theory would provide a unification of
physics. Einstein and Eddington were amongst the physicists that
attempted that unification in the 1920s and following decades. Today,
it is known that there are more fundamental fields, such as those
associated to the nuclear forces, and possibly others, and the dream
of a unified theory accounting for all fundamental fields remains,
either through an alternative gravitational theory, encompassing both
general relativity and additional fields, or through the challenging
quantisation of the gravitational field, in a framework that may
account for the remaining interactions.

\section{Black holes}
\label{in}

\subsection{Trapped regions}
The confirmation, through Eddington's observations, that light is
slightly bent by the Sun's spacetime curvature or, in a Newtonian
language, by the gravitational interaction, raises the question of how
strong this effect may become in the vicinity of a star or another
astrophysical object. To increase the magnitude of the gravitational
lens effect, the Sun should be replaced by a more compact object and the
light ray should be as tangent as possible to the object.

The most compact objects predicted by general relativity are black
holes. Strictly speaking, black holes are not material objects; they
are trapped spacetime regions. A material particle, or a light ray,
that falls into the trapped region cannot escape from there. The
trapped region is in an eternal collapsing state, where the future is
always further inside, until infinite curvature, i.e., a singularity,
is attained and spacetime, as we know it, is destroyed.

The boundary of the trapped region of a black hole that is in an
equilibrium state is called event horizon. From any point outside the
event horizon it is possible to emit a light ray such that it escapes
the gravitational pull of the black hole reaching points arbitrary
far away from it.

\subsection{Photonsphere}
However, if we 
throw a beam of light rays towards the black hole, it turns out that
the capture region is more extended than the trapped region.  That is,
there exists a neighbourhood of the event horizon such that, in case
the incoming light ray enters it, it will be inexorably captured by
the black hole, ending up by falling into the trapped region.

For the simplest black holes, which are not spinning, the outer
boundary of the capture region for light rays is called the
photonsphere. These black holes are called Schwarzschild black holes,
after the German astronomer, physicist and mathematician Karl
Schwarzschild, who found and published the corresponding mathematical
solution of the Einstein field equations in 1916.

The photonsphere is an immaterial surface exterior to the event
horizon. Light rays, i.e., photons, that start precisely tangent to
this sphere will always remain on it; the corresponding trajectory is
actually a ring, called light ring. This means that the gravitational
lens effect is so strong that the photon trajectory is bent over
itself, spanning a closed curve. These light rings are unstable
trajectories though, and any perturbation, no matter how small, will
lead the photon to either fall into the event horizon or escape
towards far away from the black hole, depending on the direction of
the perturbation.

The photonsphere is the most significant signature of the remarkably
strong gravitational lens effect created by the black hole. It has
multiple phenomenological consequences. For instance, if one considers
placing a background light source, with the black hole between the
source and the observer, one would see a dark region in the sky. Such
dark region is called the black hole shadow. As an analogy, one may
consider the shadow of the Moon during a solar eclipse. Yet, while the
latter is the silhouette of the Moon's surface, the black hole shadow
is the silhouette of the photonsphere, rather than of the event
horizon. Thus, a far away light source can probe the spacetime
geometry around the black hole only up to the photonsphere.

The shadow of a Schwarzschild black hole is always a circular
disk. But, astrophysical black holes are typically spinning and are
described by a more elaborate mathematical solution of the Einstein
field equations, found by the New Zeland mathematical-physicist Roy
Kerr in 1963. These black holes are known as Kerr black holes.

Kerr black holes are characterized by their total mass and angular
momentum, the latter parameter describing their spin. When the angular
momentum vanishes, they become Schwarzschild black holes. But a
non-vanishing rotation originates significant fundamental and
phenomenological differences. For instance, the black hole should now
be envisaged as a squashed sphere, or oblate spheroid, rather than a
round sphere, keeping nonetheless a well defined equator and a
north-south symmetry. This behaviour is in agreement with the
intuitive effect of rotation on an elastic body, such as the Earth.

The boundary of the light capturing region of the Kerr black
hole is also different from its non-spinning limit, and it is not,
typically, called a photonsphere. To understand the corresponding
concept, start by considering light rays on the black hole equatorial
plane at the threshold of the light capturing region. These orbits are
circular, hence light rings, but they depend on the relative motion
with respect to the black hole spin.  In particular, the light ring is
closer to the black hole if the photon co-rotates the black hole and
the light ring is further from the black hole if the photon
counter-rotates the black hole.

\begin{figure*}[h]
\centering
\includegraphics[scale=1.4]{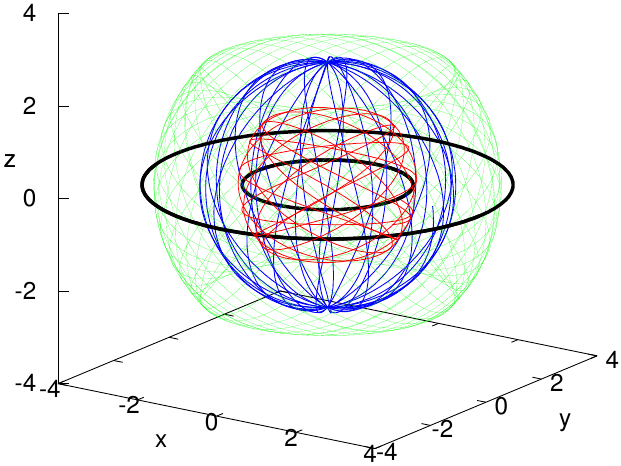}
\caption{The two light rings and some spherical photon orbits around a
Kerr black hole. The larger light ring and the green spherical orbit
are in counter-rotation with the black hole. The smaller light ring
and the red spherical orbit are in co-rotation with
the black hole.  The blue orbit has zero angular momentum.
}
\label{orbits}
\end{figure*}

Outside the equatorial plane there are also photons orbiting a black
hole, but their orbits are not confined to a plane. The black hole
rotation drags spacetime, making the non-equatorial initial orbital
plane precess. Consequently, the light ray spans a spherical section,
within a minimal and maximal latitude, symmetric to one another. There
is a continuum of such spherical orbits, interpolating between the two
light rings, see Fig.~\ref{orbits}. Starting from the exterior light
ring, counter-rotating the black hole, there are spherical orbits that
span an increasingly larger latitude range until a certain orbit spans
the whole sphere. This orbit has vanishing angular momentum. Starting
from there, another sequence of orbits, now in co-rotation with the
black hole, span a decreasing latitude range, until they degenerate
into the smaller, co-rotating, light ring.

The Kerr spherical photon orbits, including the light rings, define
the threshold for capture of incoming light rays, and, as such, the
shadow of a Kerr black hole. This shadow ceases to be, in general a
perfectly round disk and now depends on how much the black hole is
rotating and on the observation angle, relative to the equatorial
plane.  For instance, observed from the equatorial plane, a black hole
with upwards spin vector, exhibits a shadow that varies between a
circular disk, when the rotation vanishes, and a D-shape, when the
rotation is maximal. The latter shadow was first computed by the
American physicists James Bardeen in 1973.

Reciprocally, the observation of the shadow of an astrophysical black
hole can, in principle, allow the determination of how fast it is
spinning. Or even test if the observed black hole is of Kerr type or
something unexpected, which could unveil new physics. This type of
observation is not a mirage. It was carried out by the Event Horizon
Telescope, an international collaboration involving eight radio
telescopes around the world. In April 2019, the collaboration
announced it had observed the supermassive black hole at the centre of
the M87 galaxy, a black hole with an estimated 6 billion solar masses,
and an event horizon size of roughly the Solar system size, showing
for the first time the shadow of a black hole lit by the astrophysical
environment surrounding it. These 2019 results confirmed the
theoretical predictions of general
relativity. Consequently, 100 years after the observation of a small
light deflections by Eddington, we are observing the strong deflection
and lensing of light due to a black hole.

\subsection{Where are the black holes?}
Black holes were found by Oppenheimer and Snyder in a remarkable
theoretical work in 1939. The subsequent decades witnessed impressive
theoretical developments, led by the American physicist John Wheeler
and the British physicists Roger Penrose and
Stephen Hawking. On the observational side, progress as been equally
remarkable. It is estimated that of the order of ten million black
holes with a few solar masses exist in our Galaxy. Moreover,
observations suggest that all, or almost all, galaxies harbour a
central supermassive black hole. A new population of black holes with
the mass of a few dozens of solar masses was recently unveiled through
gravitational wave observations. Speculations have been put forward
that mini black holes or even black holes with masses of fundamental
particles may exist, but these have no observational support, at the
present.

\section{Gravitational waves}

\subsection{Spacetime distortions}
What happens when a star or a black hole is accelerated, or when two
stars or two black holes collide? If we use the principle, from
special relativity, that a maximal propagation speed exists for
information, one must conclude that gravity also propagates with a
finite speed. As such, when a star is accelerated, gravity falls
behind and takes some time to catch up. The equations of general
relativity show that the mathematical laws ruling this adjustment are
similar to the Maxwell equations describing electromagnetic waves
and the waves in the gravitational case also travel precisely at
the speed of light.  These waves are called gravitational
waves. Gravitational waves are propagating tidal effects or, in the
geometrical language, perturbations in the curved spacetime
traveling at the speed of light. They were predicted by Einstein in
1916.  In the same work Einstein also made the first attempt at
studying the emission of gravitational waves, but only in 1918 he
correctly reported that a varying quadrupole is necessary for such
emission. In 1922 Eddington gave the correct formula with 
a 1/2 factor that was missing in Einstein's calculations.
The mathematical
manipulation of gravitational waves is difficult, and the physical
interpretation of the results requires knowledge and insight. In the
early days of the subject, some computations suggested 
the existence of gravitational
waves with speed greater than light. Eddington showed these were
merely oscillations in the coordinate-system and the ``only speed of
propagation relevant to them is the speed of thought",
a phrase that became famous, 
demonstrating promptly that the
physical gravitational waves indeed traveled at the speed of
light. The theory of gravitational waves was subsequently developed
by many, notably the British physicist Herman Bondi, in the 1960s, and
the American physicist Kip Thorne, in the 1970s and subsequent
decades.

\begin{figure*}[ht]
\centering
\includegraphics[scale=0.7]{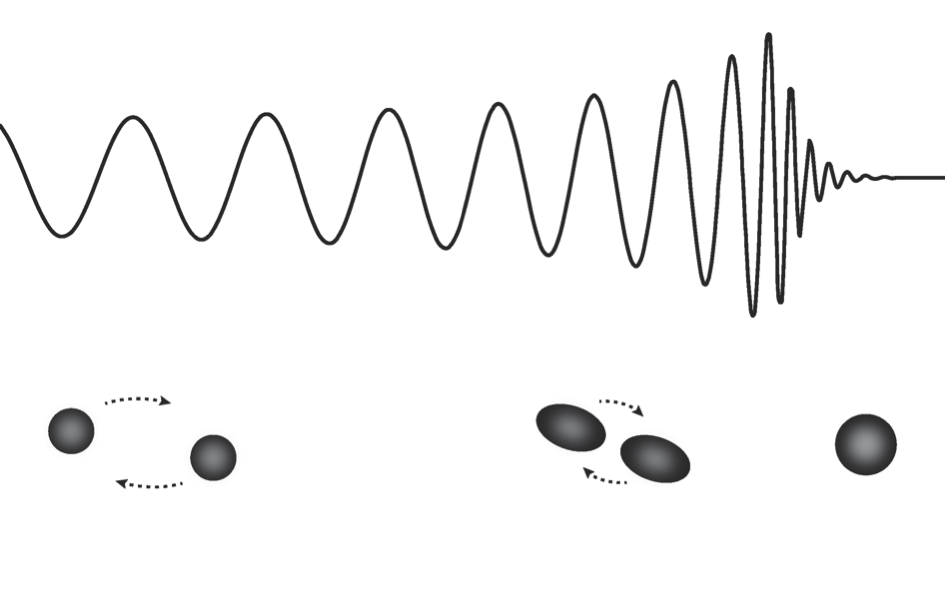}
\caption{Top panel: Representation of the waveform as a function of
time of a gravitational wave event due to the collision of two black
holes in a binary system.  The waveform provides both the amplitude
and the frequency of the wave. Bottom panel: Illustration of the
physical process that originates the gravitational wave,
namely, the inspiral,
merger, and ringdown phases of a binary black hole system.}
\label{GW}
\end{figure*}

Let us suppose that gravitational waves are
generated by a binary black hole
system, where the two black holes orbit one another somewhere in the
Universe. Those waves propagate through the cosmos at the speed of
light and eventually are detected on the Earth. The top panel of
Fig.~\ref{GW} shows, in arbitrary units, the amplitude and frequency
of the gravitational wave detected as a function of time. The wave's
amplitude is directly related with the spacetime distortion measured
relatively to a reference which is 
Minkowski spacetime. The amplitude of the wave can also be
interpreted as the magnitude of the tidal force at a given point as a
function of time. The amplitude depends on the distance the wave has
traveled, it is inversely proportional to such distance. The frequency
of the wave is twice the orbital frequency of the binary system that
generated it.

As Fig.~\ref{GW} shows, the frequency increases with time. The reason
is simple. Gravitational waves carry energy and consequently the
binary system is continuously loosing energy. Energy conservation
requires then that the black holes are approaching one another, so
that the energy carried by the gravitational waves is sourced by the
potential energy of the system. As the black holes approach, their
orbital frequency increases as time goes by and so does the frequency
carried by the waves. It can be observed
that at a given time the amplitude reaches a maximum, which
corresponds to the collision between the black holes, and thereafter
the
amplitude falls rapidly: the collision produces a bigger
black hole that rapidly relaxes to equilibrium, stopping all
gravitational wave emission. An illustration of the physical process
producing the gravitational waveform is provided in the bottom panel
of Fig.~\ref{GW}.

\subsection{The detection of gravitational waves}

The increase in orbital frequency in a binary system due to
gravitational wave emission is visible in some astrophysical
binaries. The first such observation is due to two American
physicists, Russell Hulse and Joseph Taylor, who discovered in 1974 a
binary system of two neutron stars named PSR B1913+16. This binary
system has a period of around 7 hours and an orbital radius of the
order of the Sun's diameter. It is a relativistic system. Hulse and
Taylor followed the orbit during several years, and showed that the
orbital period was decreasing at a rate of 76 microseconds per year,
with the orbital radius decreasing about 3 meters per year, in perfect
agreement with general relativity. This was the first time that
gravitational waves were detected, albeit indirectly, through the
effect of radiation damping of an orbital system. Hulse and Taylor
were awarded the 1993 Nobel physics prize for this discovery.

How can one improve what Hulse and Taylor have done and have a direct
measurement of a gravitational waves through its interaction with
matter? Since such a wave is a traveling tidal force, the detection of
gravitational waves follows the principles of tidal forces, i.e., one
should look for relative motion between two particles caused by the
passing wave, with the amplitude and frequency of the wave giving an
indication of the relative force undergone by the two particles, see
Fig.~\ref{GW}.  For instance, the gravitational wave emitted by a
black hole binary has an amplitude that decreases with the distance
travelled, and when passing in the solar system, the separation
between the Earth and the Sun, that can be considered as particles for
this effect, will undergo changes described in Fig.~\ref{GW}.  Note,
however, that when it arrives in the solar system the amplitude of the
wave is very small and the separation between the Earth and the Sun
suffers a change no larger than one thousandth of the width of a human
hair.

Despite the seemingly impossible task, Joseph Weber started the direct
search for gravitational waves in the 1960s. This project would last
half a century and involve thousands of scientists until the first
results would be obtained. The original detector design, an aluminium
bar whose length would vary if hit by a gravitational wave was
unsuccessful, but it was an important precursor of today's most
advanced detectors such as LIGO (Laser Interferometry
Gravitational-Wave Observatory). The LIGO detectors are built in two
sites in the American states of Louisiana and Washington. The working
principle of LIGO is simple. Two mirrors are hanging at the end of two
perpendicular arms with a common origin, in which a laser beam
circulates. When a weak, but with appropriate amplitude and frequency,
gravitational wave reaches the Earth, the length of the arms varies
infinitesimally. This variation of length is measured by the
interference properties of the laser beam which is being reflected at
the two mirrors. In this apparatus, the mirrors are the particles
whose relative motion is triggered by the wave. Measuring the way in
which the length of the arms varies in time, one may infer the
properties of the sources that generated those waves, and with several
detectors one can also know from where the waves came. Actually,
decades of theoretical and technological efforts were required, in one
of the most complex and longest pursuits for an experimental
confirmation of a theoretical prediction.  The Earth shakes
constantly, the laser exerts pressure on the mirrors, all of this with
many other factors is cause of noise that hinders a clean detection
of gravitational waves.

\subsection{GW150914 and the future}
The search was a success story. On September 14th 2015, the scientist
on duty looked at the screen and saw that the mirrors were moving
precisely as in Fig.~\ref{GW}! This gravitational wave event, named
with the date on which it was detected, GW150914, was the first observation of the
collision between two black holes, each with about 30 solar masses,
the collision taking place at a cosmological distance of 1200 million
light years.

The Nobel Physics prize
of 2017 was awarded to the direct detection of gravitational waves
in the names of Ray Weiss, the experimentalist
that established how one can overcome the noise problem
in such a detector, Kip Thorne, the
theoretician behind the detection, and Barry Barish, the coordinater
of the LIGO collaboration with hundreds of physicists.

Afterwards, LIGO and the European detector Virgo, situated in Pisa,
saw many more events of black hole collisions, besides another spectacular
collision of two neutron stars also observed in all frequencies of the
electromagnetic spectrum.  It is expected that LIGO will start
observing one or more collisions per week when it achieves design
sensitivity.

Due to the Earth's continuous shaking, i.e., seismic noise, it is
difficult to carry out Earth based observations of gravitational waves
in the frequency range below 10 Hz. This range precisely matches that
range where gravitational waves are generated from events involving
supermassive black holes, like the ones at galactic centres, are
expected. To tackle this problem, the European Space Agency will
construct the space-based detector LISA (Laser Interferometer Space
Antena). It is also expected that the primordial universe generated
gravitational waves due to small perturbations. Inflationary theories
of the early universe make precise predictions about these
perturbations and their gravitational wave spectrum. It is possible
that LISA may see such waves, which would lead to an extraordinary
advance in our knowledge of the early universe.

\section{Conclusions}

Since the creation of general relativity by Einstein in 1915 and its
confirmation via the deflection of light observations in the 1919
eclipse by Eddington and collaborators, we have witnessed a constant
development and remarkable discoveries. Black holes were predicted in
1939 and confirmed by observations spanning the whole range of the
electromagnetic spectrum in subsequent decades. It is known today that
all, or almost all, galaxies have a central black hole. In 2015,
gravitational waves were confirmed by direct observation.  The first
detected waves came from two black holes in a binary system at a
cosmological distance that collided generating an enormous quantity of
energy in the form of gravitational waves, an infinitesimal part of
which passed by the Earth.

But there is more, much more, to do and understand. For instance,
when, where and how black holes formed? How did they grow? How do they
influence galaxy development? The black holes we are observing are the
ones predicted by Einstein's general relativity? It is expected that
more detailed observations of black holes will bring some
answers. What is dark matter and how does it interact with the
standard model matter? The only interaction known for dark matter is
gravitational. Is it possible to see the effects of dark matter
through gravitational waves?

General relativity cannot be the final gravitational theory. The
interior of black holes is singular and at the singularity spacetime
itself is destroyed. How can we observe the interior of a black hole?
What is a spacetime singularity made of? We do not have answers, yet,
but science has always surprised us with what we thought was beyond
our reach.

\newpage

\noindent
{\bf Credits for the figures:}
\noindent
Figure 1 was taken from  P.~V.~P.~Cunha, C.~A.~R.~Herdeiro, and E.~Radu,
Physical Review D 96, 024039 (2017).
\noindent
Figure 2 is a representation of the event GW150914 detected by
LIGO, see Abbott et al  (LIGO Scientific Collaboration and Virgo Collaboration),
Physical Review Letters 116, 061102 (2016).

\vskip 0.3cm
\noindent
{\bf Portuguese version of this article:}
\noindent
This article has been first published in Portuguese, see
J.~P.~S.~Lemos, C.~A.~R.~Herdeiro,  V.~Cardoso, ``Einstein e
Eddington e as consequências da relatividade geral: Buracos negros e
ondas gravitacionais'', Gazeta de Física {\bf 42(2)}, 36 (2019).  It
is part of a special number of Gazeta de Física dedicated to
Einstein, Eddington, and the Eclipse, marking the celebrations of the
100 years of light deflection, see A. J. S. Fitas, P. Crawford, and
J. P. S. Lemos (editors), {\it Einstein, Eddington, Eclipse} (Número
especial dedicado à exposição E3 - Einstein Eddington e o Eclipse,
Gazeta de Física, Lisbon, 2019).

\vskip 0.3cm
\noindent
{\bf Note:}
C. A. R. Herdeiro is now at
Centre for Research and Development in Mathematics and
Applications - CIDMA, Departamento de Matemática,
Universidade de Aveiro,
Campus de Santiago, 3810-183 Aveiro, Portugal, email: herdeiro@ua.pt.

\vskip 0.3cm
\noindent
{\bf Acknowledgements:}
We acknowledge Funda\c c\~ao para a Ci\^encia e Tecnologia (FCT)
Portugal for financial support through Grant~No.~UID/FIS/00099/2019.

\end{document}